# Atomistic-Continuum Hybrid Simulation of Heat Transfer between Argon Flow and Copper Plates

**Yijin Mao, Yuwen Zhang and C.L. Chen**

Department of Mechanical and Aerospace Engineering
University of Missouri
Columbia, Missouri, 65211
Email: zhangyu@missouri.edu

**Abstract**

A simulation work aiming to study heat transfer coefficient between argon fluid flow and copper plate is carried out based on atomistic-continuum hybrid method. Navier-Stokes equations for continuum domain are solved through the Pressure Implicit with Splitting of Operators (PISO) algorithm, and the atom evolution in molecular domain is solved through the Verlet algorithm. The solver is validated by solving Couette flow and heat conduction problems. With both momentum and energy coupling method applied, simulations on convection of argon flows between two parallel plates are performed. The top plate is kept as a constant velocity and has higher temperature, while the lower one, which is modeled with FCC copper lattices, is also fixed but has lower temperature. It is found that, heat transfer between argon fluid flow and copper plate in this situation is much higher than that at macroscopic when the flow is fully developed.

**Nomenclature**

| | |
|---|---|
| $A$ | surface area expose to argon flow, m$^2$ |
| $C$ | volumetric heat capacity, J/m$^3$ K |
| $c_p$ | specific heat, J/kg-K |
| $D$ | characteristic length, m |
| $e$ | internal energy, J |
| $\mathbf{f}_i$ | force acting on i$^{th}$ atom, N |
| $h$ | heat transfer coefficient, W/m$^2$K |
| $k$ | thermal conductivity, W/m K |
| $k_B$ | Boltzmann constant, $1.38\times10^{-23}$ J/K |
| $m_i$ | mass of atom i, kg |
| $M_J$ | mass of liquid argon of J$^{th}$ control volume, kg |
| $n$ | coupling interval |
| $N$ | number of atoms |
| $Nu$ | Nusselt number |
| $p$ | pressure, Pa |
| $\mathbf{r}_{ij}$ | vector from pointing from atom j to i |
| $r_c$ | cutoff distance, m |
| $\mathbf{r}_i$ | position vector of atom i |
| $t$ | time, s |
| $T$ | temperature, K |
| $\mathbf{u}$ | average velocity of total atoms in one control volume, m/s |
| **U** | velocity, m/s |
| **U**$_J$ | velocity of liquid argon of $J^{th}$ control volume, m/s |
| v$_i$ | velocity of atom i, m/s |
| V | potential energy (or Volume of simulation box), J |
| *Q* | Total energy passes through the copper wall, J |

**Greek Symbols**

| | |
|---|---|
| *α* | damping factor |
| δ*t*$^P$ | time-step in molecular dynamics simulation |
| *ε* | minimum potential energy |
| *ρ* | density |
| *σ* | depth of potential well |
| *υ* | viscosity |

**Introduction**

Fluid dynamics and heat transfer behaviors in micro fluidics have drawn intensive attentions in the last two decades due to the rapid development of MEMS/NEMS and many other micromechanics applications [1-3]. A better scientific understanding on fundamental mechanism at such small scale will definitely bring a favorable impact on in the foreseeable future. For example, an improved understanding of thermal conductivity from atomic point of view reveal the causes leading to thermal damage of the computer chip which is supposed to be thermally safe under the conventional Fourier law. It is often found that some experimentally measured parameters under micro- spatial/temporal scale, such as heat transfer coefficient at solid-fluid interface and thermal conductivity at solid-solid interface, dramatically disagree with the ones predicted through conventional theory for macro-scale, due to size effect [4]. In order to better understanding the heat transfer mechanism in micro-/nano-scale, numerical simulation is an effective and promising alternative approach.

It is well known that the widely applied three conservation laws can resolve problems for macroscopic scale. However, due to the break-down of continuum assumption, it is also understood that an advanced theory should be developed to remedy the subsistent disadvantage of current conservation laws based simulation tools. Thus, classical molecular dynamics simulations are emerging as another powerful tool to provide detailed information on phonon scattering, which further can be used to calculate corresponding thermal properties through certain formula, such as Green-Kubo formulism. A faithful representation of dynamic system should be spatially and temporally large and long enough [4]. As a result, such level of simulation is far beyond the most advanced super computer simulation capability.

As a compromise and meanwhile to take full advantages of both sides, a hybrid simulation scheme that solve three conservation equations in larger domain while resolve atomic trajectory in smaller domain could be a promising approach at the present time. In fact, several effective hybrid simulation methods have been developed to study these particular phenomena caused by the size effect; these methods include atomic finite element method (AFEM) [5], atomistic-smooth particle method [6], and atomistic-finite volume method[7]. The AFEM has advantage of high computational efficiency for solid state problems, and the smooth-particle method is a simulation technique that is still under development [8], which also suffers issues from unclear physical meaning at boundary. A literature survey indicates that atomistic-finite volume method is the most popular hybrid approach among those similar hybrid schemes [9-17].

Since many problems are tangled with fluid flow, finite volume method based hybrid scheme are widely adopted. Yasuda and Yamamoto carried out the hybrid simulations for some element flow of Lennard-Jones liquids and demonstrated the validity of this hybrid simulation scheme. Barsky[18] use this scheme to study dynamic of a single tethered polymer in a solvent. A series of conformational properties of the polymer for various shear rates are obtained. Yasuda, and Yamamoto [19] have demonstrated that the hybrid simulation of molecular dynamic and CFD is able to reach a good consistency as long as the mesh size and time step of CFD domain is not too large comparing to the system size and sample during in MD simulation. Wang and He [20] have developed a dynamic coupling model for a hybrid atomistic-continuum in micro- and nano-fluidics. However, most of them primarily emphasize on momentum coupling, while only a few account for energy exchanging at the coupling region. Liu et al. [21] developed computer codes that coupled continuum-atomistic simulator and conducted simulation on heat transfer in micro-/nano-flows. Sun et al. [22] have developed molecular dynamics-continuum hybrid scheme and studied condensation of gas flow in a micro-channel.

An atomic-continuum hybrid simulation of convective heat transfer between argon and two parallel copper plates is carried out. The continuum domain is solved using OpenFOAM [23], while the molecular domain is described using LAMMPS[24]; there is an overlap domain for data exchange between the continuum and molecular domains. This approach is similar to of the recently released CFDEM library [25], but with significant extension. A solver that is fully capable of solving continuum domain and atomic domain through multi-processor super computer system is developed under PISO (Pressure Implicit with Splitting of Operators) solving procedure. It should be mentioned that a reduced unit system is employed in this simulation model. The characteristic dimensions of length, energy, and mass are depth of Lenard-Jones potential well, the minimum value of the potential $\varepsilon$, and the mass m of argon atom, for argon. An asterisk (*) will be used to represent the reduced quantities such as the reduced length $r^* = r/\sigma$, and the reduced temperature $T^* = Tk_B/\varepsilon$, the reduced pressure $P^* = P\sigma/\varepsilon$, the reduced mass density $\rho^* = \rho\sigma^3/m$, the reduced time $t^* = t\,(\varepsilon/m/\sigma^2)^{1/2}$. All the results will be converted back to the dimensional form for discussion.

**Physical Models and Methods**
**Domain Decomposition**
In order to solve the problem, a three dimensional model that involve liquid argon and solid copper wall is created. Since the entire computational domain will be resolved with finite volume method (FVM) and atom-based molecular dynamics, the computational domain is decomposed into three regions: continuum region (C), atomic region (P), and overlapped region (O) between C and P regions (see Figure 1). In the C region, the fluid flow will be solved with FVM based on classical Navier-Stokes equations, while the P region will be resolved with classical molecular dynamics with an appropriate potential function. A set of coupling schemes, which will be described in the following section, is adopted to guarantee that momentum and heat flux through specified control layers laying at both the top and bottom of O region, namely, P→C and C→P regions, are continuous.

**Governing equations for continuum region (C)**
Since the density change during the process is negligible, the argon flow in the continuum region is considered to be incompressible. Meanwhile, the flow is considered to be unsteady, thus any minor effects, such as velocity fluctuation altered through atomic domain during small time-step simulation can be conveyed back and further affects the behaviors in the C region. The governing equation for C region can be expressed as follows:

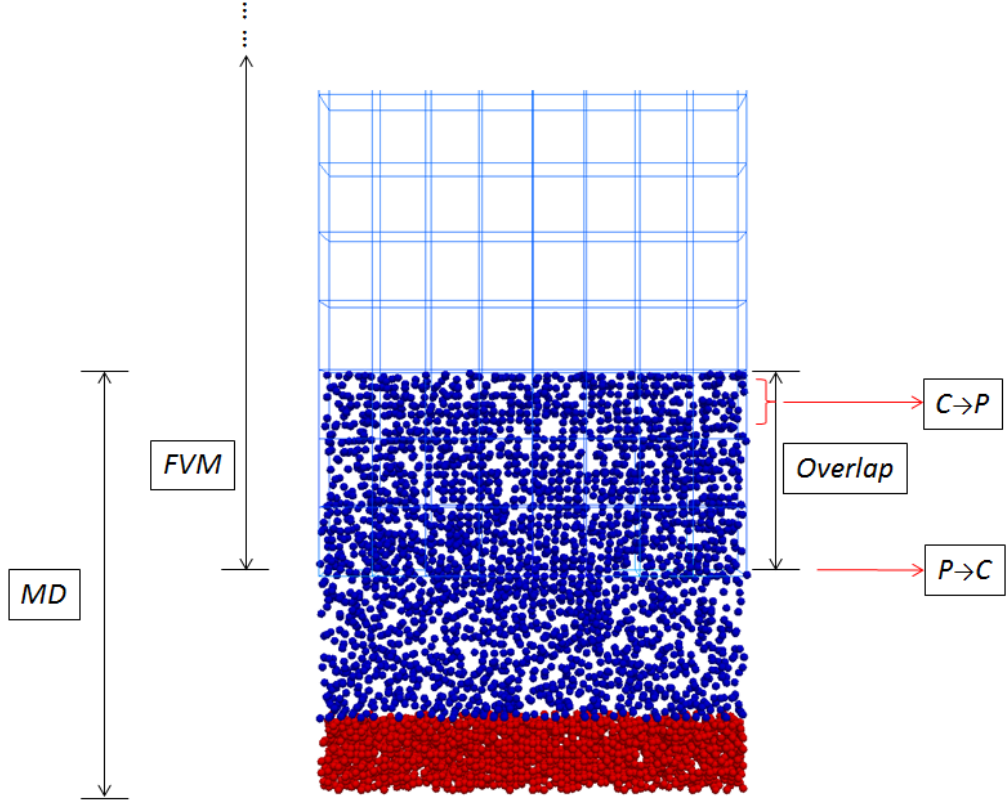
Figure 1 Schematic of domain decomposition

Continuity equation:
$$\nabla \cdot \mathbf{U} = 0 \qquad (1)$$

Momentum equation:
$$\frac{\partial \mathbf{U}}{\partial t} + (\mathbf{U} \cdot \nabla)\mathbf{U} = \nu \nabla^2 \mathbf{U} - \frac{1}{\rho}\nabla p \qquad (2)$$

Energy equation:
$$\frac{\partial T}{\partial t} + \mathbf{U} \cdot \nabla T = \frac{k}{\rho C_v} \nabla^2 T \qquad (3)$$

It is important to point out that thermal conductivity, dynamic viscosity, and specific heat are treated as constants since the temperature difference within the C domain is relatively small.

**Potential functions for the atomic region (P)**

In the atomic domain, classical molecular dynamics simulation is adopted to resolve the trajectory evolution of argon atoms. An important assumption for classical molecular dynamics states that the potential energy is a function that only depends on atomic positions. Therefore, an appropriate potential function is essential to describe the interaction among atoms and to govern the dynamic behaviors of atoms in the simulation box. For the interaction between argon atoms, the following commonly accepted modified Lennard-Jones potential, which is able to avoid abrupt energy decrease near cutoff distance, will be used:

$$V_{LJ}(r_{ij}) = 4\varepsilon \left[ \left(\frac{\sigma}{|r_{ij}|}\right)^{12} - \left(\frac{\sigma}{|r_{ij}|}\right)^{6} - \left(\frac{\sigma}{r_c}\right)^{12} + \left(\frac{\sigma}{r_c}\right)^{6} \right] \qquad (4)$$

where $\varepsilon$ is minimum potential energy and $\sigma$ is depth of argon Lennard-Jones potential well, $r_{ij}$ is vector connecting atoms i$^{th}$ and j$^{th}$, and $r_c$ is cutoff radius which usually chosen as a length of $3.5\sigma$.

For the potential function of metallic atoms, the Lennard-Jones type function is not well-defined because the contribution from free electrons, which play a key role in thermal transport in most metal, is oversimplified. In fact, as stated in [26], the contribution of free electrons to the thermal conductivity is approximately 100 times more than that of the lattices. Embedded-atom method (EAM) [27] generated potential is a special potential that consider free electron effect on interatomic interaction between ions by introducing an electron density function. And it is fitted to the following mathematical form based on experimental data,

$$V_{EAM}(r_{ij}) = F_i\left(\sum_{j\neq i} f_j(r_{ij})\right) + \frac{1}{2}\sum_{j\neq i}\varphi(r_{ij}) \qquad (5)$$

where the first term on the right hand side accounts for the electrons contribution due to interaction between electron gas and nuclei. The second term $\varphi$ account for the two-body interaction between the nuclei. Since the classical molecular dynamics assume that electron position relative to nuclei is rigidly fixed (Born-Oppenheimer approximation), which lead to a huge simplification to avoid cumbersome computation for the molecular system, the atomic system can be easily described by solving a set of Newtonian mechanics equations:

$$m_i \frac{d^2 \mathbf{r}_i}{dt^2} = -\sum_{i\neq j} \frac{\partial V(\mathbf{r}_{ij})}{\partial \mathbf{r}_{ij}} \qquad (6)$$

where $m_i$ is atomic mass of $i^{th}$ atom, $r$ is the position of atom $i$, and V is corresponding potential energy.

**Coupling (Overlap) region**
**Momentum coupling**
In the overlap region, a control layer (C→P) is constructed to achieve the momentum consistencies between the atomic and continuum domains. The momentum consistency implies that the mean momentum in the molecular region should be equal to the instantaneous macroscopic momentum from continuum region, i.e.:

$$\langle m_i \mathbf{v}_i \rangle = M_J \mathbf{U}_J \qquad (7)$$

where the left hand side is spatial average of momentum, and the right hand side represents the momentum in the specific control volume.

In order to achieve this momentum consistency, an external force that is proportional to the momentum difference at the same location between the one that from continuum level and the other one from molecular domain will exert on atoms within the control layer (see C→P layer in Figure 1). In fact, the velocity in each control volume can be explicitly obtained by solving momentum equations. Another velocity from molecular dynamics within the same control volume can be estimated by:

$$\mathbf{u} = \left\langle \sum_{j=1}^{N^{P-C}} m_j \dot{\mathbf{r}}_j \bigg/ \sum_{j=1}^{N^{P-C}} m_j \right\rangle \qquad (8)$$

Thus, the force that will act on each atom in the C→P region can be estimated through its acceleration:

$$\ddot{\mathbf{r}}_i = \frac{1}{\left(\frac{n}{2}+1\right)}\left(\mathbf{u}_{CFD} - \left\langle \sum_{j=1}^{N^{P-C}} m_j \dot{\mathbf{r}}_j \bigg/ \sum_{j=1}^{N^{P-C}} m_j \right\rangle_{n\delta t^P}\right) + \left(\frac{\mathbf{f}_i}{m_i} - \left\langle \frac{1}{N^{P-C}}\sum_{j=1}^{N^{P-C}} \frac{\mathbf{f}_j}{m_j} \right\rangle_{n\delta t^P}\right) \qquad (9)$$

where the first term reflects the velocity difference and the second term is to adjust the strength of molecules motion. Number *n* represent coupling interval between two different domains, $\delta t^P$ is the time-step in molecular dynamic simulation, $N^{P\text{-}C}$ is number molecules within control volume, and the force *f* is the one derived from potential function. In addition, the second term represents the inherent fluctuation in molecular system to accelerate or decelerate molecules. The summation of this term within the control volume is zero [9]. It should be pointed out that even the second term is only expressed as $f_i/m$, it also obtains good results as reported in some researchers' work [10, 20]. At the same time, the velocity values estimated at the local cell that involve bottom boundary will be assigned as boundary condition of the continuum domain, as shown as P→C in Figure 1.

**Temperature coupling**

Similar to the momentum coupling, the temperature field from atomic domain should also match that from the continuum domain. In order to synchronize temperature, each local cell that has atoms in the control layer is connected to a thermal reservoir that holds the temperature of collocated control volume in the continuum domain. Meanwhile, the temperature estimated through kinetic theory will be assigned to continuum domain, through P→C layer. The temperature from molecular domain is estimated by:

$$T = \frac{2}{3k_B N^{P\text{-}C}} \left\langle \sum_{j=1}^{N^{P\text{-}C}} \frac{1}{2} m \left( \dot{\mathbf{r}}_j - \mathbf{u} \right)^2 \right\rangle \tag{10}$$

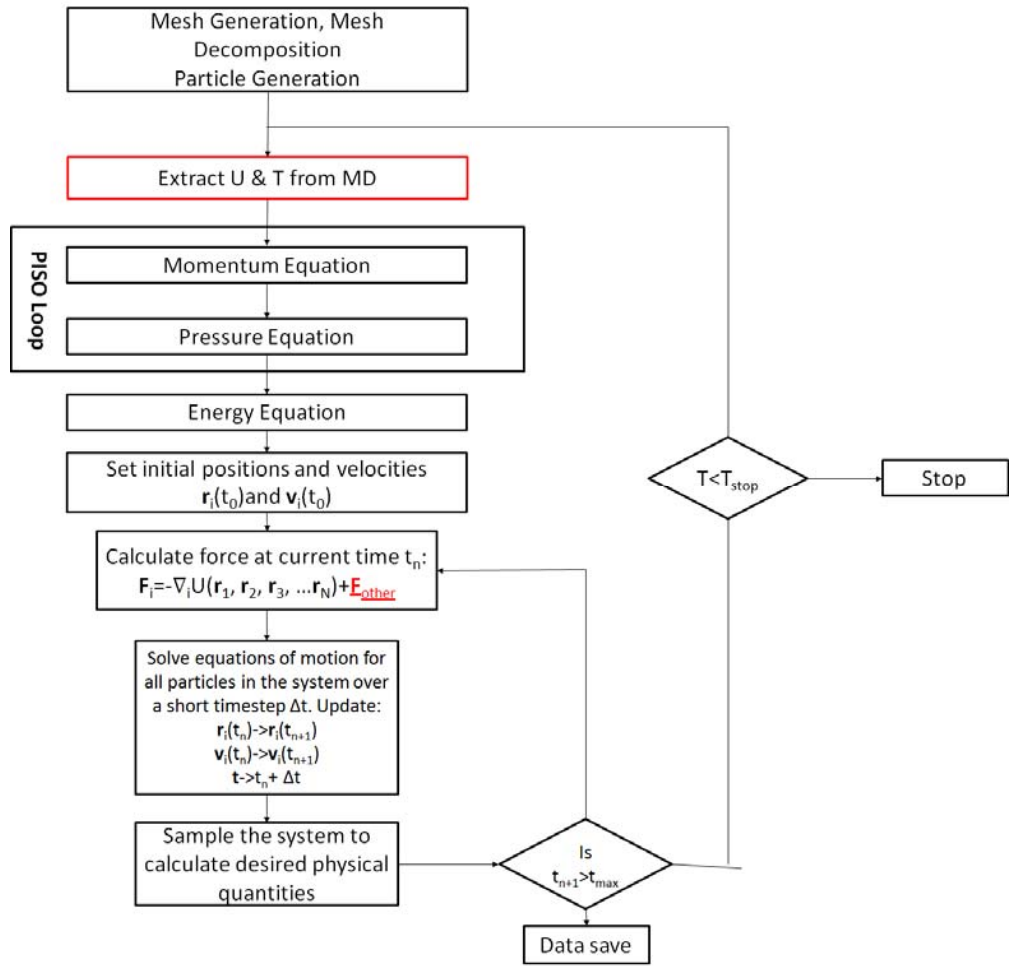

Figure 2 Computation flow chart

where the velocity **u**, which is considered as the bulk velocity of a group of molecules located within one control volume, is obtained from the continuum level computation.

The above thermal reservoir that will be connected to the control volume in the control layer is realized by Langevin method[28]. The acceleration of atom i is calculated from:

$$\ddot{\mathbf{r}}_i = \alpha \left( \dot{\mathbf{r}}_i - \frac{1}{N^{P-C}} \sum_{j=1}^{N^{P-C}} \dot{\mathbf{r}}_j \right) + \frac{\mathbf{f}_i}{m} + \frac{\mathbf{F}}{m} \qquad (11)$$

where the first term on the right hand side represents the thermal fluctuation, $\mathbf{f}_i$ is the force estimated through potential function, $\mathbf{F}$ is a random force vector that satisfy Gaussian distribution with mean value of 0 and standard deviation of $(2\alpha k_B T/\delta t^P)^{1/2}$, where $\alpha$ is damping factor and $T$ is target temperature[28]. In this work, the velocity difference is assumed to be immediately eliminated so that the damping factor is set to be 1.0.

Figure 2 shows the computational flowchart which integrate MD evolution into CFD solver based on the solver pisoFoam. Momentum and pressure equations are solved through the PISO algorithm, and temperature field will be resolved afterwards. Classical molecular dynamics simulation will be performed for certain FVM time-step (coupling interval) after reading atoms' velocity and position successively. Then the results from MD will be extracted and transferred back to FVM domain, which will be solved again. The entire simulation ends when the iteration reaches to the total simulation time.

**Results and Discussions**
**Code Validation**
In order to validate the implemented schemes, a two-dimensional computational domain which includes both continuum and atomic part is created, as shown in Figure 3. For both CFD and MD region, the length and height of the domain is 46.9$\sigma$ by 46.9$\sigma$. It is assumed that the entire domain is filled with liquid argon. For the continuum domain, the fluid has a constant density of $0.83m/\sigma^3$, kinematic viscosity of $1.144\sigma^2/\tau$, thermal diffusion of $0.598\sigma^2/\tau$, where the Lennard-Jones parameter $\sigma$, $\varepsilon$, mass $m$, Boltzmann constant $k_B$ are unit, and reduced time $t$ is defined as $(\varepsilon/m/\sigma^2)^{1/2}$. The grids configuration is 10 by 10 in x and y direction. And the molecular domain is modeled with 7,390 Lennard-Jones type atoms. It should be pointed out that the top edge of molecular domain is considered to be an argon atoms composed wall in order to mimic the interaction between atoms in MD domain and phantom in CFD domain; of course, this treatment can also avoid atom loss. The simulation time step is 0.001 $\tau$. In order to validate reliability of momentum and energy coupling, two simple cases, which include pure fluid dynamics and pure heat conduction problem, are solved and compared with published analytical solution.

Couette flow that is driven by a constant moving top wall is a classical problem that is often used for code validation. In order to validate momentum coupling scheme, only the momentum coupling operation will be applied to the overlap domain. In other words, energy transfer procedure, which is also close related to thermal velocity of atoms, will not be activated to alter velocity exchange in this region. However, a thermal reservoir that holds a constant temperature of 1.1$k_B/\varepsilon$ will be connected to the molecular domain, such that the argon atoms will have a desired constant temperature.

Figure 3(a) shows that final configuration of this test case, where the molecular domain has both liquid atoms and solid atoms. These solid atoms are used to compose a static wall that mimic physical boundary. And the top wall velocity is set to be (1.0$\sigma/\tau$, 0, 0). Figure 3(b) shows several velocity profiles along the vertical direction at different times, which are 50$\tau$, 100$\tau$, 200$\tau$, and 2000$\tau$, respectively. It can be seen that the velocity obtained from FVM and MD simulation in the overlap region is consistent with each other at all sampled time, which directly demonstrate that the momentum scheme implement works in a desirably. In fact, the final velocity profile, which is a straight line, also agrees upon the analytical solution [20].

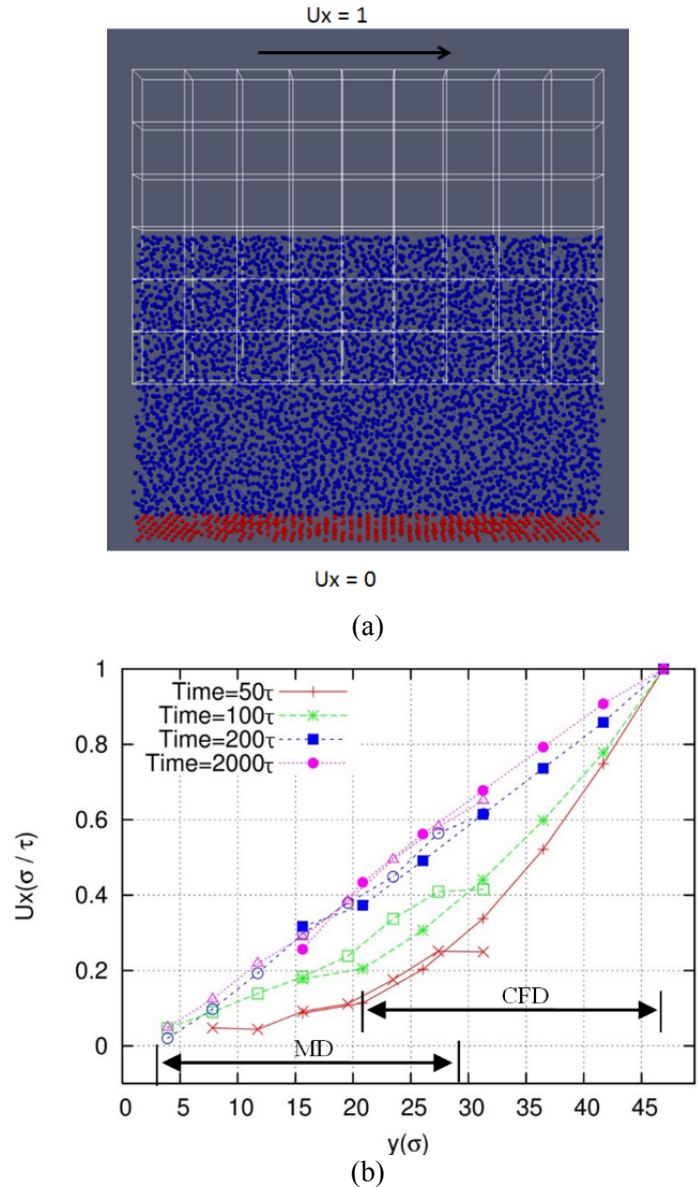

Figure 3 Couette flow: (a) configuration, (b) final velocity profiles

To test energy coupling in the overlap domain, heat conduction for both continuum domain and molecular domain is solved. The energy equation in continuum domain is solved to obtain temperature, and only energy coupling is activated through the overlap region this time. The temperature of the top wall of continuum domain is fixed at 1.5 $k_B/\varepsilon$ while the bottom wall of molecular domain is set to be 1.1 $k_B/\varepsilon$, as shown in Figure 4(a). Figure 4(b) shows temperature profiles at different sample times. It can also be seen that the temperature profile along the vertical line, at different time, is gradually being a flat straight line, which is also agree well with results from other researchers[20].

**Convection Heat Transfer in Couette Flow**
By following the hypothesis initialized by Tuckerman and Pease[29], the heat transfer coefficient may vary significantly when the size approaching to micro-scale. One observation can be done from the perspective of Nusselt number, whose definition is Nu=$hD/k$, where $h$ is convective heat transfer coefficient, $D$ is hydraulic diameter, and $k$ is thermal conductivity.

It is noticed that $h$ is possible to scaled up to thousand or millions times if the hydraulic diameter $D$ reduce to micro- or nano- size for a fully developed flow in the micro-channel, if the definition of Nu number is still valid and thermal conductivity is constant at all scales.

In this section, heat transfer coefficient $h$, between argon flow and solid copper surface will be estimated by solving momentum and energy equation simultaneously with finite volume method in continuum domain and simulating the wall-close domain with molecular dynamic approach; momentum and energy are coupled through the overlap region at the same time.

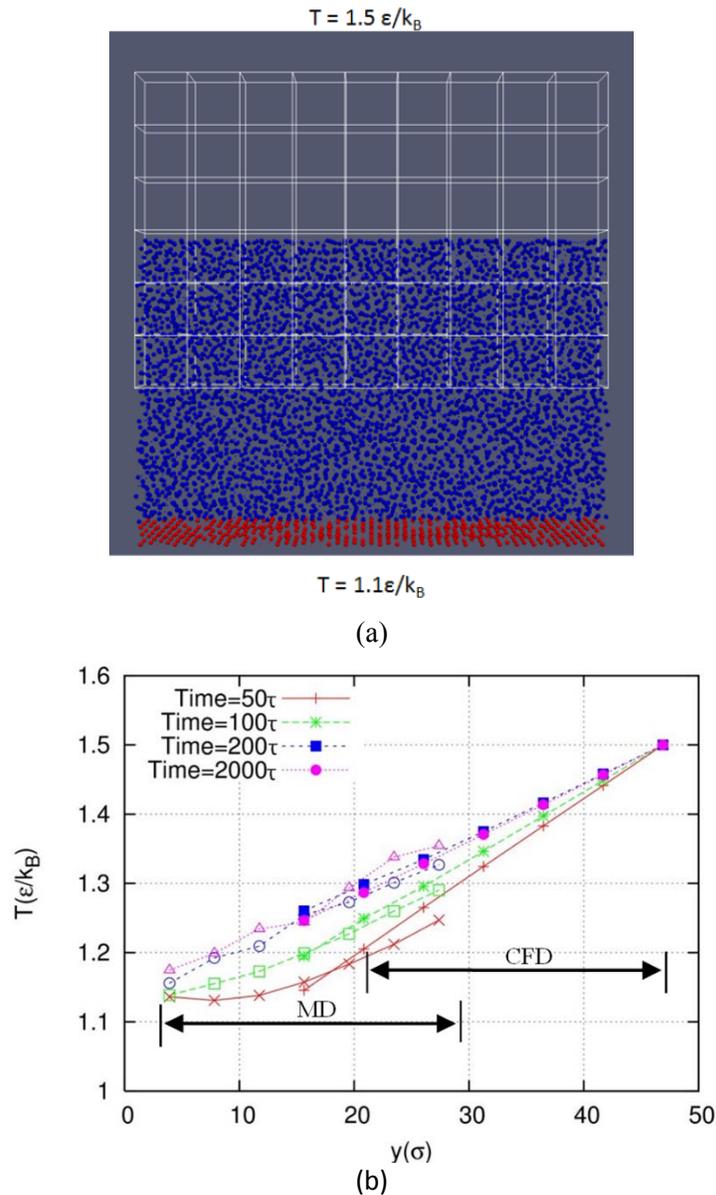

Figure 4 Heat conduction: (a) configuration, (b) final temperature profiles

A new setup is created to study convective heat transfer efficiency by determining heat flux flow across the domain. In comparison with the simple test cases, the layout of the entire computational domain is same, but the atoms in wall domain are replaced with copper atoms. In addition, EAM potential [33] that is able to account for thermal effect contributed by energy carrier of free electron(interaction between electron and phonons) at the very beginning of heating process. The interaction among argon atoms and the interaction between argon and copper atoms are both described with classical Lennard-Jones-type potential. The parameters in Lennard-Jones potential for argon-argon interaction, $\sigma_{Ar-Ar}$ and $\varepsilon_{Ar-Ar}$ are simply $\sigma$ and $\varepsilon$ [30] for argon. For interaction between argon and copper atoms, $\sigma_{Ar-Cu}$ and $\varepsilon_{Ar-Cu}$ are $0.64\sigma$ and $6.24\varepsilon$ [31], which is obtained based on geometric mixture manner[32]. There are 7,390 atoms in the atomic domain that holds the same dimension as the cases for validation. In other words, the flow density of atomic domain is consistent with the one in the continuum domain. But the bottom wall is replaced with copper plate that has different lattice constant consistent with density of $3.17\sigma^3$. For the temperature configuration, a Langevin type thermal reservoir is used to control temperature of the copper wall that holds a constant temperature of $1.1\varepsilon/k_b$. Meanwhile, the top wall of the continuum domain has a fixed temperature of $1.5\varepsilon/k_b$.

For velocity arrangement, a fixed velocity of $(1.0\sigma/\tau, 0, 0)$ is applied to the top wall. It is worth to notice that no jump condition is applied either to temperature or velocity boundary of the top and bottom wall, due to the fact that Knudsen of the system is around 0.01 [33]. For the thermal reservoirs, a damping factor, which determine the interaction frequency with thermal reservoir [24], of 2 is used, instead of 1 after trial and error, and they are connected to control volumes within control layer, such that the temperature information can be successfully transmitted to the atomic domain. The coupling interval is 100 time-steps. The time-step is set to be $0.005\ \tau$. The cases run for $4000\tau$ in total, such that the flow in the computational domain is fully developed. In order to achieve a statistically meaningful result, 10 similar cases that have almost the same configurations except for initial atomic positions and velocities are simulated. In addition, in order to measure the heat flux flowing from hot to cold end, a special region between top and bottom wall is created to compute the energy flux when the flow reach to steady state. Figure 5 shows temperature and velocity profiles along the vertical line when the flow is under steady state, where both velocity and temperature profile do not changed with time. It can be observed that the temperature and velocity at coupled zone work well as expected. Both temperature and velocity are consistent with those from the other region, within an acceptable statistical error.

Finally, the heat transfer coefficient between argon flow and copper wall is estimated through equation as following,

$$h = \frac{Q}{A\Delta t\Delta T} \qquad (12)$$

where $Q$ it the total energy passes through copper wall, $\Delta t$ is the time period that this amount of energy pass through, $\Delta T$ is the temperature difference between the copper wall and bulk temperature of argon flow, and $A$ is the surface area of the wall that expose to the argon flow. From the atomic perspective, the heat flux can be estimated through [34]:

$$q = \frac{Q}{A\Delta t} = \frac{1}{V}\left[\sum_i e_i \mathbf{v}_i - \sum_i S_i \mathbf{v}_i\right] = \frac{1}{V}\left[\sum_i e_i \mathbf{v}_i + \sum_i \left(\mathbf{f}_{ij}\cdot\mathbf{v}_j\right)\mathbf{r}_{ij}\right] \qquad (13)$$

where $e$ represents internal energy of atom, $v_i$ is velocity of atom, $\mathbf{f}_{ij}$ is pair-wise force between atoms, and $\mathbf{r}_{ij}$ is relative position between atoms i and j.

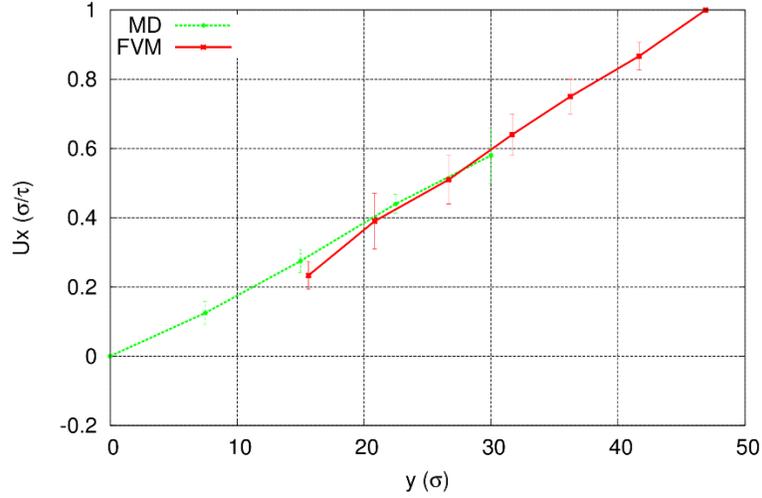

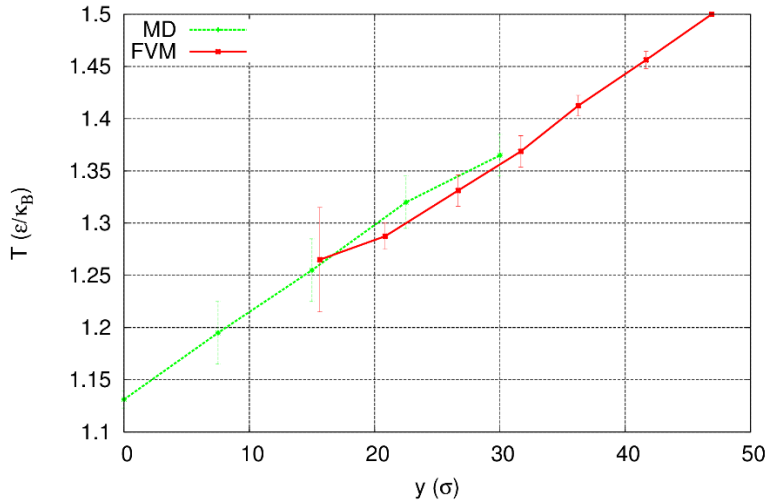

(b)
Figure 5 Velocity and Temperature profiles for Couette flow with heat transfer

Figure 6 shows the real-time variation of heat flux passing through preset region in the atomistic domain. It can be seen that the recorded heat flux fluctuated within a certain and small range. It worth to point out that the sign change of heat flux is mainly caused by large noise and relatively small temperature gradient across the entire atomistic computational domain. It is expected that the relative noise level will significantly decrease with higher temperature gradient. In this simulation work, however, it is not appropriate to impose a large temperature gradient due to a small temperature difference between melting and boiling points of the argon.

For this reason, the computational domain is limited to current domain sizes. As shown in Figure 6, a clear trend shows that the heat flux is negative though some values are above zero due to noise. The heat flux, which is $-0.086\varepsilon/\sigma^2\tau$, adopted to compute heat transfer coefficient is an averaged value over the latest period of $1000\tau$, based on 10 cases. It is found that the heat transfer coefficient between argon flow and copper plate is $0.25k_b/\sigma^2\tau$ ($1.21\times10^7$ W/m$^2$ K), which is larger but still in the same order magnitude with the one ($0.83\times10^7$ W/m$^2$ K, with Nu=1.38, $k$=96.4 mW/m K [35]) predicted based on the hypothesis in reference [36], where the expected value may increase significantly if the definition of Nusselt is assumed to be valid and thermal conductivity is constant at micro-scale.

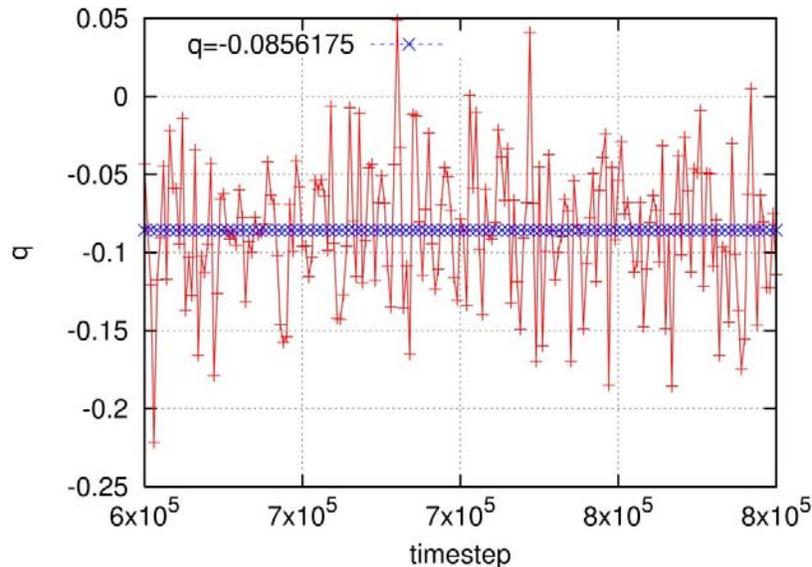

Figure 6 Variation of heat flux variation with simulation time-step

**Conclusions**

In this work, atomistic-continuum hybrid simulation of heat transfer between argon flow and copper plate is carried out, based on a general solver that is developed based on OpenFOAM and LAMMPS. The validity of this coupling scheme is tested through a Couette flow case and a heat conduction case. The solver achieves satisfactory agreement with the existing results for testing cases, which mean momentum and energy successfully meet continuous requirement at the overlap region. Finally, heat transfer coefficient between flowing argon and solid copper plate is studied within this frame work. Both velocity and temperature are coupled at the overlap region simultaneously, and reach to stable profiles in vertical direction. Then heat flux flowing across the sample region is measured in order to compute heat transfer efficiency. It is found that the heat transfer coefficient is larger but still in same order magnitude with the one predicted based on the hypothesis in reference [36]. Further investigation is desired to explore heat transfer coefficient for micro-channel with other flow medium (such as water), which has supportive experimental data, and also allow to impose high temperature to the top and bottom walls.

**Acknowledgments**

Support for this work by the National Science Foundation under grant number CBET- 1066917 and the Office of Naval Research under grant number N00014-14-1-0402 is gratefully acknowledged.